\documentstyle[epsf,12pt]{article}
\input math_macros
\newdimen\rotdimen
\def\vspec#1{\special{ps:#1}}
\def\rotstart#1{\vspec{gsave currentpoint currentpoint translate
   #1 neg exch neg exch translate}}
\def\rotfinish{\vspec{currentpoint grestore moveto}}
\def\rotl#1{\rotdimen=\ht#1\advance\rotdimen by\dp#1%
   \hbox to\rotdimen{\vbox to\wd#1{\vskip\wd#1\rotstart{270 rotate}%
   \box#1\vss}\hss}\rotfinish}%
\def\dt{\!\cdot\!}
\def\roughly#1{\raise.3ex\hbox{$#1$\kern-.75em\lower1ex\hbox{$\sim$}}}
\def\o{\over}
\def\l{\left}
\def\r{\right}
\def\dr{$\overline{\it DR}$~}

\begin{document}
\begin{titlepage}
\begin{center}
December, 1993 \hfill       JHU-TIPAC-930030\\
\hfill PURD-TH-93-13\\
\hfill hep-ph/9312248\\
\vskip .7 in
{\large \bf Radiative Corrections to Neutralino and Chargino\\
            Masses in the Minimal Supersymmetric Model}
\vskip .3 in
           \vskip 0.5 cm
      {\bf Damien Pierce}\\
      {\it Department of Physics and Astronomy\\
          The Johns Hopkins University\\
         Baltimore, Maryland\ \ 21218\\}

           \vskip 0.5 cm
      {\bf Aris Papadopoulos}\\
      {\it Physics Department\\
           Purdue University\\
           West Lafayette, Indiana\ \ 47907\\}

\end{center}
\vskip 0.4 in
\begin{abstract}
We determine the neutralino and chargino masses in the MSSM at one-loop.
We perform a Feynman diagram calculation in the on-shell renormalization
scheme, including quark/squark and lepton/slepton loops.
We find generically the corrections are of order 6\%.
For a 20 GeV neutralino the corrections can be larger than 20\%. The
corrections change the region of
$\mu,\ M_2,\ \tan\beta$ parameter space which is ruled out by LEP data.
We demonstrate that, e.g., for a given $\mu$ and $\tan\beta$
the lower limit on the parameter $M_2$ can shift by 20 GeV.
\end{abstract}
\end{titlepage}
\renewcommand{\thepage}{\arabic{page}}
\setcounter{page}{1}
\newbox\rotbox

\section{Introduction}
The Minimal Supersymmetric Model (MSSM) provides a concise theoretical
framework in which supersymmetry is realized in a simple and consistent
manner. The parameter space of the MSSM is somewhat constrained by
present experimental data. As is well known, the region of parameter
space ruled out by LEP experiments due to Higgs boson searches is
dramatically altered when radiative corrections are taken into account.
The large shift in the Higgs boson mass can be approximated by
considering the diagram in Fig.(1a). The correction to the squared mass is
approximately
\begin{equation}
\Delta m_h^2 \simeq 3\dt4{\lambda_t^2\o16\pi^2}m_t^2
\ln\l({\tilde{m}_t^2\o m_t^2}\r)\qquad\Longrightarrow
\qquad{\Delta m_h^2\o m_h^2}\simeq 60\%\label{delta-mh}
\end{equation}
\begin{figure}[h]
\epsfxsize=8cm
\epsffile[0 510 280 650]{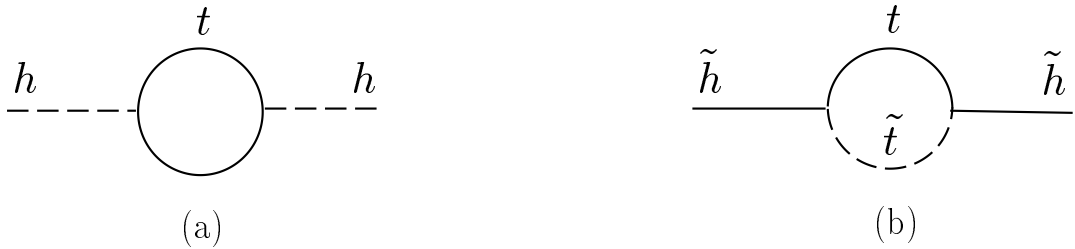}
\caption[f5]{(a) Top quark contribution to the Higgs boson mass.
(b) Top/stop contribution to the Higgsino mass.}
\end{figure}
(We consider $m_h=M_Z$ at tree level. Here and henceforth
we set the top quark mass to 150 GeV, and the squark mass to 1 TeV.)
In Eq.(\ref{delta-mh})
there is a factor of 3 from color, a factor of 4 from a Dirac trace,
two factors of  the top quark Yukawa coupling $\lambda_t$ from the
vertices, a loop factor $1\o16\pi^2$,
two factors of $m_t$ from mass insertions, and a leading logarithm from
the logarithmically divergent integral. This correction is especially
important phenomenologically at LEP II, where if one did not take
into account radiative corrections the MSSM would be ruled out by a negative
Higgs boson search \cite{HBosonI}. In fact, after inclusion of these
corrections, the $m_A,\ \tan\beta$ parameter space is only mildly
constrained \cite{HBosonII}.

Motivated by large corrections in the Higgs
boson sector, one can ask if we can also expect large corrections
for the Higgs super-partners. The correction to the Higgsino mass can be
estimated by inspecting the diagram in Fig.(1b). In this case we find
$${\Delta M_{\tilde{h}}\o M_{\tilde{h}}} \simeq 3 {\lambda_t^2\o16\pi^2}
\ln\l({\tilde{m}_t^2\o m_t^2}\r)\simeq 5\%.$$
(We consider $M_{\tilde{h}}$=90 GeV.) The factor of 12 enhancement in the
Higgs scalar case compared to the Higgsino case is due to the Dirac trace and
the factor of three enhancement from $m_t^2/M_Z^2$. Hence we can
expect Higgsino mass corrections to be mild compared to the Higgs boson
mass corrections. A similar analysis for the gaugino mass correction
also yields an estimate for the correction of 5\%.

In the MSSM the supersymmetric partners of the $W$, $Z$, and photon
(the gauginos) mix with the partners of the Higgs bosons
(the Higgsinos) to give the mass eigenstates. The
spectrum then consists of two charged states (the charginos) and four neutral
states (the neutralinos). The charginos are denoted $\chi_1^+,\ \chi_2^+$,
while the neutralinos are $\chi_1^0,\ \chi_2^0,\ \chi_3^0,$ and
$\chi_4^0$. They are arranged in order of increasing mass, so that
$\chi_1^+\ (\chi_1^0)$ is the lightest chargino (neutralino).
The explicit formulas for the neutralino masses and eigenvectors
at tree level can be found in Refs. \cite{tree_masses}.

The parameter space which determines the chargino and neutralino masses
at tree level includes the supersymmetric Higgs mass parameter
$\mu$, the soft--supersymmetry breaking $U(1)_Y$ and $SU(2)_L$
gaugino masses $M_1$ and $M_2$, and the ratio of Higgs boson vacuum
expectation values $\tan\beta=v_2/v_1$. In this paper we assume the GUT
relation among gaugino masses, so we set
$M_1 = {5\o3}\tan^2\theta_{_W} M_2$. Thus we will typically examine the
corrections in the $\mu$,\ $M_2$ plane for fixed values of $\tan\beta$.

In the next section we describe the formalism necessary to describe the
radiative corrections. In section 3 we discuss the results, and in the
last section we give our conclusions.

\section{Formalism for radiative corrections}
In this section we outline our renormalization scheme.
The chargino and neutralino masses are determined at tree level by the
bare parameters $x_{i_b}=\l\{M_{W_b}^2,\ M_{Z_b}^2,\ M_{1_b},\ M_{2_b},
\ \mu_{_b},\ \beta_{_b}\r\}$.
At one-loop we must choose a renormalization prescription for each of these
parameters which determines the renormalized parameter $x_{i_r}$
and the shift $\delta x_i$, where $$x_{i_b} = x_{i_r} + \delta x_i.$$
We choose the renormalization prescription for the parameters $M_{W_b}^2$ and
$M_{Z_b}^2$ so that at one-loop the parameters $M_{W_r}$ and $M_{Z_r}$
are the poles of the $W$ and $Z$ propagators. Hence
$$\delta M_W^2 = {\rm Re~} \Pi_{WW}^T(M_W^2),\qquad\quad
\delta M_Z^2 = {\rm Re~} \Pi_{ZZ}^T(M_Z^2)$$
where $\Pi^T$ denotes the transverse part of the boson propagator.
Formulas for these gauge boson self-energies can be found in Ref.\cite{HZZ}.
A convenient renormalization prescription for the remaining parameters
$M_1,\ M_2,\ \mu,$ and $\beta$ is the \dr prescription wherein the shifts
$\delta x_i$ are purely ``infinite", i.e. proportional to
$\l(1/\epsilon + \ln4\pi - \gamma_E\r)$. For these parameters
we choose the \dr renormalization scale to be $Q^2=M_Z^2$.
We have previously determined the shift $\delta\beta$ while studying radiative
corrections in the Higgs boson sector \cite{HZZ}. To determine the remaining
shifts $\delta M_1,\ \delta M_2,$ and $\delta\mu$ we first discuss the
physical masses.

The physical on-shell chargino and neutralino masses,
defined as the poles of the propagators, are given by \cite{Aoki}
\addtocounter{equation}{1}
$$
{M_{\chi^+_i}}_{\rm phys} =  M_{\chi^+_{ir}} + \delta M_{\chi^+_i}
                -{\rm Re}\left(\Sigma^+_{1_{ii}}(M_{\chi^+_i}^2)+
              M_{\chi^+_i} \Sigma^+_{\gamma_{ii}}(M_{\chi^+_i}^2)\right)
\eqno{(\theequation{\rm a})}$$
$$
{M_{\chi^0_i}}_{\rm phys} =  M_{\chi^0_{ir}} + \delta M_{\chi^0_i}
                -{\rm Re}\left(\Sigma^0_{1_{ii}}(M_{\chi^0_i}^2)+
              M_{\chi^0_i} \Sigma^0_{\gamma_{ii}}(M_{\chi^0_i}^2)\right)
\eqno{(\theequation{\rm b})}\label{physical_mass}
$$
where we have substituted $M_{\chi_{ir}} + \delta M_{\chi_i}$
for the bare mass $M_{\chi_{ib}}$, and the $\Sigma_{ij}$'s are form
factors of the one-loop fermion inverse propagator $K_{ij}$
\begin{equation}
iK_{ij}=\l(\rlap/p -M_{\chi_{ib}}\r)\delta_{ij}
+ \Sigma_{1_{ij}}+\Sigma_{5_{ij}}\gamma_5
+\Sigma_{\gamma_{ij}}\rlap /p +\Sigma_{5\gamma_{ij}}\rlap /p\gamma_5
\end{equation}
The bare chargino masses $ {M_{\chi^+_{ib}}}$ are related to bare
parameters $M_{2_b},\ \mu_{_b},\ \beta_{_b}$ and $M_{W_b}$ by the equations
\addtocounter{equation}{1}
$$M^2_{\chi^+_{1b}}+M^2_{\chi^+_{2b}}= M^2_{2_b}+\mu_{_b}^2+2M_{W_b}^2
\eqno{(\theequation{\rm a})}$$
$$M^2_{\chi^+_{1b}} M^2_{\chi^+_{2b}}
 =\left( M_{2_b}\mu_{_b}-M_{W_b}^2 \sin(2\beta_{_b})\right)^2
\eqno{(\theequation{\rm b})}$$
whereas the bare neutralino masses are the absolute values of the
eigenvalues of the bare mass matrix
\begin{equation}
{\bf Y}=\left(
\begin{array}{cccc}
M_{1_b}       &     0     &
            -M_{Z_b}c_{\beta_b}s_{W_b} & M_{Z_b}s_{\beta_b}s_{W_b}\\
0          &     M_{2_b}   &
             M_{Z_b}c_{\beta_b}c_{W_b}&-M_{Z_b}s_{\beta_b}c_{W_b} \\
-M_{Z_b}c_{\beta_b}s_{W_b} & M_{Z_b}c_{\beta_b}c_{W_b}& 0 & -\mu_{_b}\\
M_{Z_b}s_{\beta_b}s_{W_b}   & -M_{Z_b}s_{\beta_b}c_{W_b}& -\mu_{_b} & 0
\end{array}\right)
\end{equation}
where $s_\beta\ (c_\beta)$ denotes $\sin\beta\ (\cos\beta)$,
$c_W$ denotes $\cos\theta_W=M_W/M_Z$, and $s_W=\sin\theta_W$.
We introduce the matrix {\bf N} which diagonalizes the neutralino mass matrix,
\begin{equation}
{\bf N}^{\star} {\bf Y} {\bf N}^{-1} = {\rm Diag}(M_{\chi_{ib}^0}).
\end{equation}

The shifts in the input parameters $\delta x_i$ induce shifts in the bare
masses $M_{\chi_b}$. From Eqs.(4) the shifts $\delta M_{\chi^+_i}$
are explicitly related to $\delta M_2,\ \delta \mu,\ \delta\beta$ and
$\delta M_W^2$ by the following equations
\addtocounter{equation}{1}
$$M_{2}\,\delta\! M_2+\mu\,\delta\!\mu = \l[M_{\chi^+_{1}}\,\delta\!
M_{\chi^+_1}+M_{\chi^+_{2}}\,\delta\! M_{\chi^+_2}-\delta\! M^2_W\r]_\infty
\eqno{(\theequation{\rm a})}$$
$$M_{2}\,\delta\!\mu+\mu\,\delta\! M_2 =\Biggl[\frac{ M_{\chi^+_1}
M_{\chi^+_2}}
{ M_2\mu-M_{W}^2\sin(2\beta) }\left(M_{\chi^+_1}\,\delta\! M_{\chi^+_2}
+M_{\chi^+_2}\,\delta\! M_{\chi^+_1} \right)
\eqno{(\theequation{\rm b})}$$
$$\qquad\qquad\qquad\qquad\qquad+\;\;
2M^2_{W}\cos(2\beta)\,\delta\!\beta+\sin(2\beta)
\,\delta\! M^2_W\Biggr]_\infty$$
where the subscript $\infty$ denotes the ``infinite" part,
while the shifts $\delta\! M_{\chi^0_i}$ are given by
\begin{equation}
\delta\! M_{\chi^0_i}= \left({\bf N}^{\star} {\bf \delta Y}{\bf N}^{-1}
                                                       \right)_{ii}
\end{equation}
where ${\bf \delta Y}$ is the matrix whose elements are the shifts of
the elements of the matrix ${\bf Y}$.
We note that
\begin{equation}
\sum_{i=1}^4\eta_{i}\delta M_{\chi^0_i}=
\sum_{i=1}^4({\bf \delta Y})_{ii}=\delta\!M_1+\delta\!M_2.
\label{trace}
\end{equation}
Here $\eta_i=\pm 1$ depending on whether $M_{\chi_i^0}$ is equal to $+$ or $-$
the corresponding eigenvalue of the matrix ${\bf Y}$. (See
Refs.\cite{Haber&Kane, Gunion&Haber} for a detailed discussion of
this technical point).

Clearly equation (\ref{physical_mass}a) determines the ``infinite" part
of the chargino mass shift to be
$$ \l[\delta M_{\chi^+_i}\r]_\infty
= \l[\Sigma^+_{1_{ii}}(M_{\chi_i^+}^2)
+M_{\chi_i^+}\Sigma^+_{\gamma_{ii}}(M_{\chi_i^+}^2)\r]_\infty $$
and hence we can determine $\delta M_2$ and $\delta\mu$ from
Eqs.(7). Equation(\ref{physical_mass}b)
determines analogously the ``infinite" part of $\delta M_{\chi_i^0}$,
whereby we determine $\delta M_1$ from the trace equation, Eq.(\ref{trace}),
$$\delta M_1 = -\delta M_2
+ \l[\sum_{i=1}^4\eta_i\delta M_{\chi_i^0}\r]_\infty$$
Having determined all the shifts in the underlying parameters, we then
find the radiatively corrected chargino and neutralino on-shell masses,
given by Eq.(\ref{physical_mass}), are indeed finite. It is nontrivial to check
that all individual neutralino masses are free of divergences.

Finally we give the explicit formulas for the chargino and neutralino
self-energy form factors which are necessary for the above calculation.
The top quark contribution to the chargino form factors is given by
\addtocounter{equation}{1}
$$\Sigma^{+}_{1_{ii}}(p^2)= {N_c\o16\pi^2}
m_t\left( |a^+_{t\tilde{b}i}|^2-|b^+_{t\tilde{b}i}|^2\right)
                      B_0(p^2,m_t^2,\tilde{m}_b^2)
\eqno{(\theequation{\rm a})}\label{chip}$$
$$\Sigma^{+}_{\gamma_{ii}}(p^2)={N_c\o16\pi^2}
\left( |a^+_{t\tilde{b}i}|^2+|b^+_{t\tilde{b}i}|^2\right)
                      B_1(p^2,m_t^2,\tilde{m}_b^2)
\eqno{(\theequation{\rm b})}$$
Here $N_c$ is the number of colors, and $B_0$ and $B_1$ are the
standard integrals that appear in one-loop
two-point function calculations. Explicit formulas may be found in
Ref.\cite{Kniehl}.
The chargino--top quark--bottom squark coupling $g_{t\tilde{b}\chi^+_i}$
is parameterized by $a^+_{t\tilde{b}i}$ and $b^+_{t\tilde{b}i}$ as
$g_{t\tilde{b}\chi^+_i}= a^+_{t\tilde{b}i} + b^+_{t\tilde{b}i}\gamma_5$.
In the couplings $a^+_{t\tilde{b}i},\ b^+_{t\tilde{b}i}$
and in Eqs.(10) we suppress the squark index.
We implicitly sum over the bottom squarks $\tilde{b}_1$ and $\tilde{b}_2$
in Eqs.(10). It is straight forward to generalize Eqs.(10) to include
the contributions from the other quarks and leptons.
\newpage

For the neutralinos we have similar formulas. The top quark
contribution to the neutralino form factors is
\addtocounter{equation}{1}
$$
\Sigma^0_{1_{ii}}(p^2)= {N_c\o8\pi^2}
m_t\left( |a^0_{t\tilde{t}i}|^2-|b^0_{t\tilde{t}i}|^2\right)
                            B_{0}(p^2,m_t^{2},\tilde{m}_t^2)
\eqno{(\theequation{\rm a})}$$
$$
\Sigma^0_{\gamma_{ii}}(p^2)= {N_c\o8\pi^2}
\left( |a^0_{t\tilde{t}i}|^2+|b^0_{t\tilde{t}i}|^2\right)
                            B_{1}(p^2,m_t^{2},\tilde{m}_t^2)
\eqno{(\theequation{\rm b})}$$
Here again we suppress the squark index. Implicit in Eqs.(11) is a
sum over top squarks $\tilde{t}_1$ and $\tilde{t_2}$.

As a check on the calculation, we find that the $\beta$-constants derived from
Eqs.(2b,11b) in the limit of a pure bino, wino, or Higgsino eigenstate
agree with the standard RGE equations for the parameters $M_1,\ M_2$,
and $\mu$ \cite{Falck}. Additionally, we checked that the
correction Eq.(11a) in the appropriate limit agrees with the results
of Ref.\cite{Barbieri} obtained for a massless photino.
(As previously pointed out in Ref.\cite{Alabama} there is a
$\tilde{t}_1,\ \tilde{t}_2$ top squark mixing angle factor $\sin2\theta_t$
missing from Eq.(4) of
Ref.\cite{Barbieri}, which is unity in the context of Ref.\cite{Barbieri}.)
The chargino and neutralino couplings can be found in Ref.\cite{Gunion&Haber}.

\section{Results}
At tree level the neutralino and chargino masses are invariant under
$\mu\rightarrow -\mu,\ M_2\rightarrow-M_2$ (and $M_1\rightarrow-M_1$).
At one-loop level this invariance is violated by typically less than 0.1\%. It
is weakly broken only because the squark masses are not invariant under
$\mu \rightarrow -\mu$. Hence, we shall show results only for $M_2>0$. In the
results shown here we set the soft supersymmetry breaking squark and slepton
mass parameters $M_Q=M_U=M_D=M_E=M_L=1$ TeV, we set the the squark and slepton
`$A$-term' parameters $A=$200 GeV, and the top quark mass is
set to 150 GeV.

In Figs.(2a-f) we show contours of $\chi$ masses  at
tree level (dashed lines) and one-loop level (solid lines)
in the $\mu$, $M_2$ plane with $\tan\beta=2$. Note that results shown for the
lightest chargino and neutralino in Figs.(2a,c) are qualitatively similar, and
the contours for the heaviest chargino and neutralino in Figs.(2b,f) are
quantitatively similar, both at tree level and at one-loop.

For the heaviest neutralino ($\chi_4^0$) and chargino ($\chi_2^+$)
the radiative corrections yield positive shifts in the mass
$\Delta M_\chi$ of $\sim2$ GeV for $\mu$ and $|M_2|\simeq$ 50 GeV increasing
to 30 GeV for  $|M_2|\simeq 500$ GeV. We show in Fig.(3a) contours of the
correction $\Delta M_{\chi_2^+}$ in the $\mu,\ M_2$ plane with $\tan\beta=4$.
Figure (3a) is nearly identical to the corresponding figure for the
heaviest neutralino.
In the region $|M_2|>|\mu|$ the heaviest neutralino and chargino are
predominantly wino and they couple to the matter particles via the $SU(2)$
gauge coupling. Hence the


\begin{figure}[t]
\epsfxsize=6.25cm
\setbox\rotbox=\hbox{\epsffile[115 100 375 725]{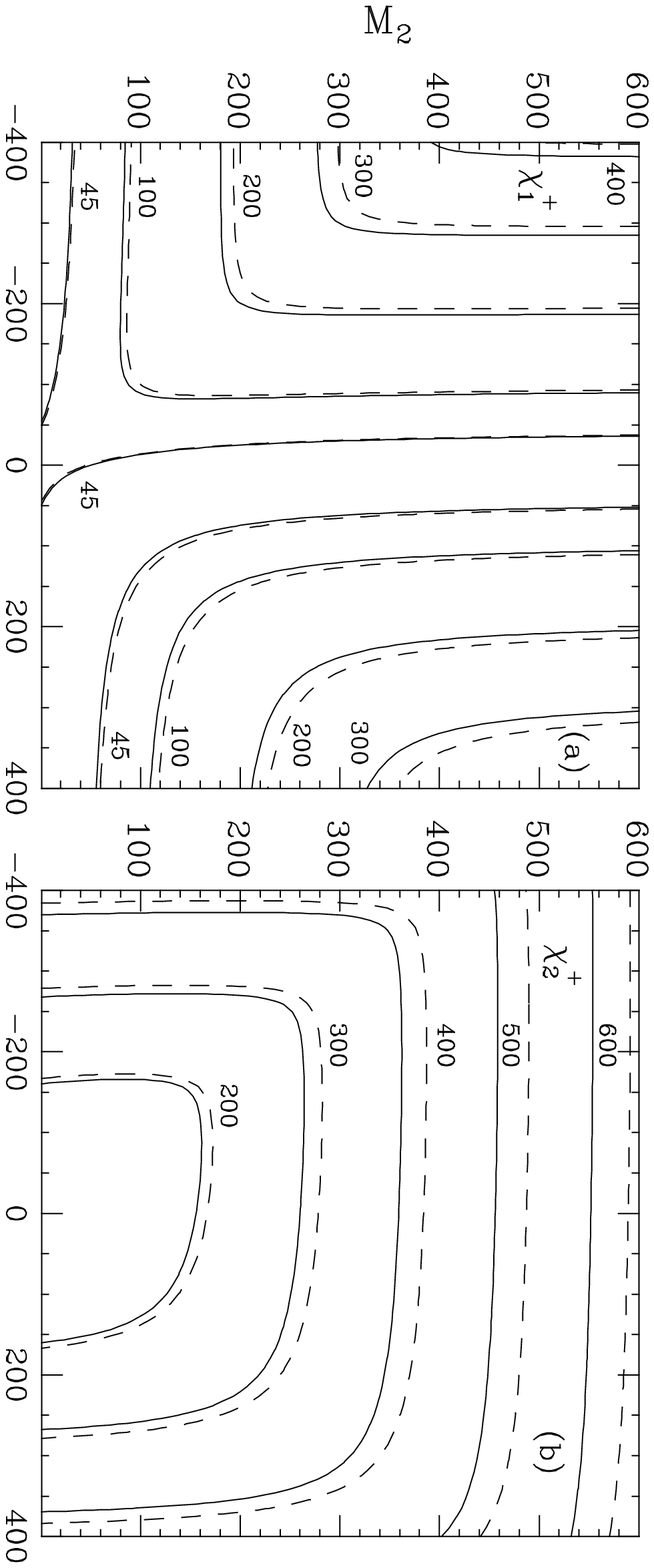}}\rotl\rotbox
\epsfxsize=6.25cm
\setbox\rotbox=\hbox{\epsffile[137 100 397 725]{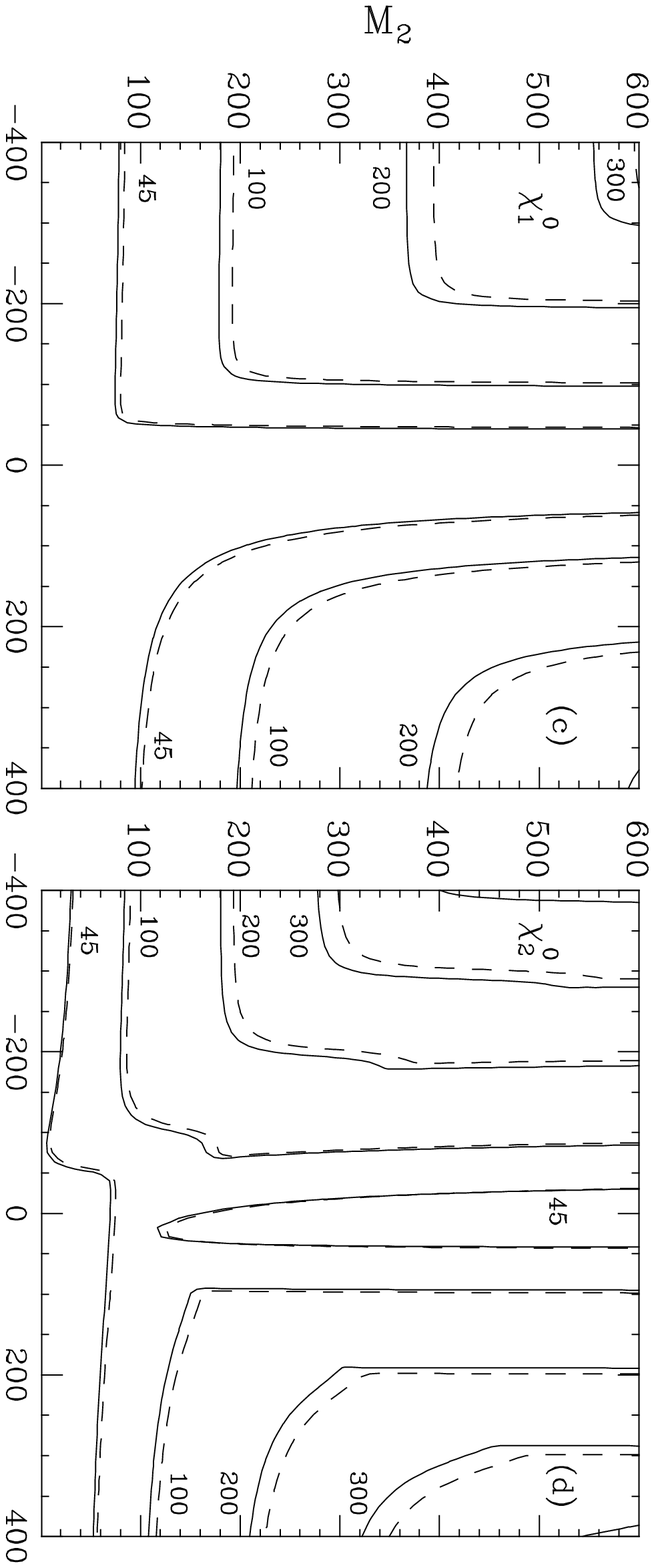}}\rotl\rotbox
\epsfxsize=6.25cm
\setbox\rotbox=\hbox{\epsffile[159 100 419 725]{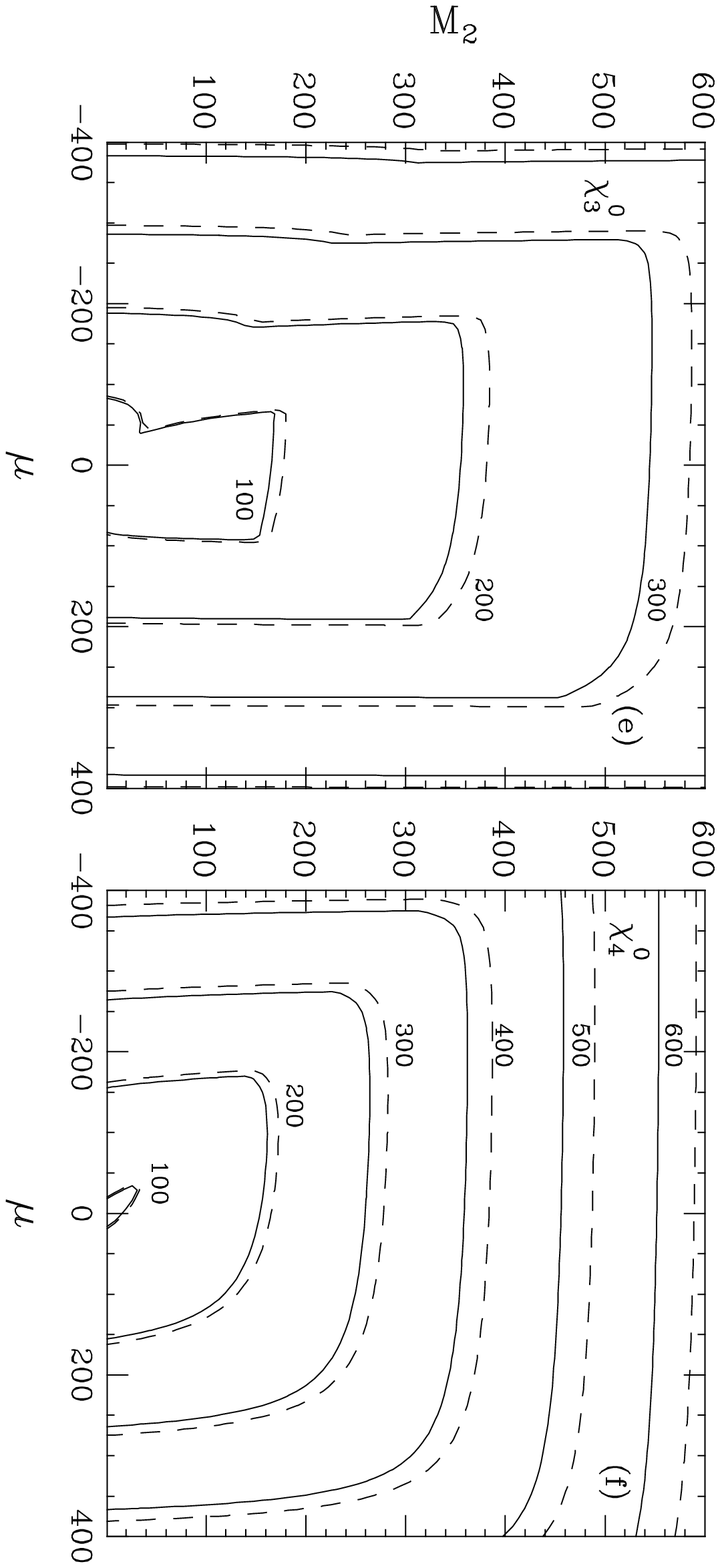}}\rotl\rotbox
\caption[f5]{Contours of chargino and neutralino masses in the $\mu,\ M_2$
plane at tree (dashed lines) and one-loop (solid) level with $\tan\beta=2$. The
axes and contours are labeled in GeV units.}
\end{figure}
\clearpage


\begin{figure}[tb]
\epsfxsize=7cm
\setbox\rotbox=\hbox{\epsffile[150 100 440 725]{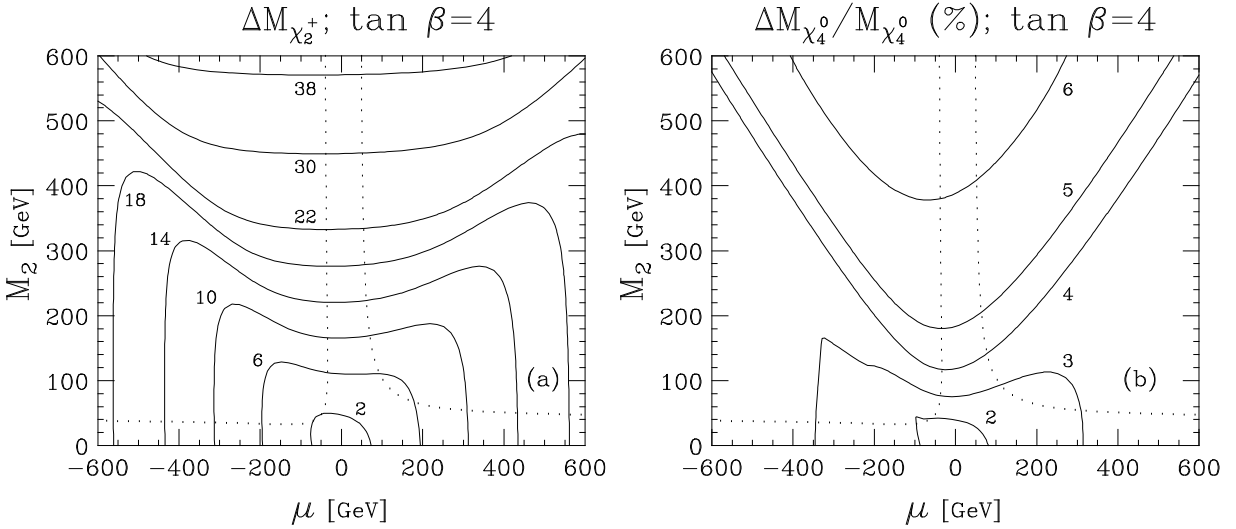}}\rotl\rotbox
\caption[f5]{(a) Contours of the correction $\Delta M_{\chi^+_2}$
to the heavy chargino mass in the $\mu,\ M_2$ plane, at $\tan\beta=4$.
The contours are labeled in GeV. (b) The percent change of the heavy neutralino
mass $M_{\chi_4^0}$ in the $\mu,\ M_2$ plane at $\tan\beta=4$.
In both figures the dotted lines are the contours of $M_{\chi^+_1}=45$ GeV.}
\end{figure}


\begin{figure}[htb]
\epsfxsize=7cm
\setbox\rotbox=\hbox{\epsffile[150 100 440 725]{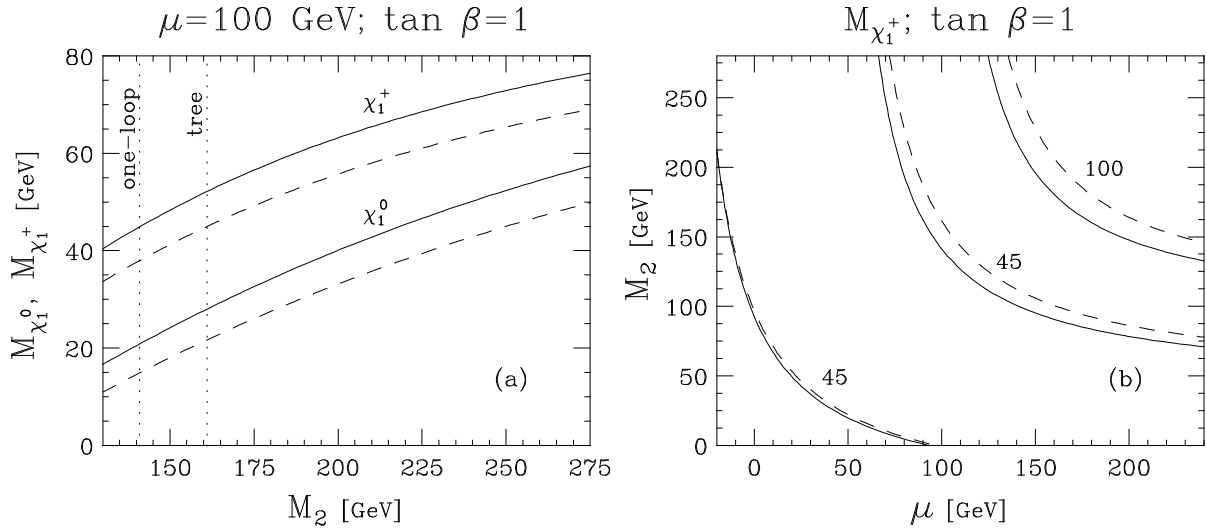}}\rotl\rotbox
\caption[f5]{(a) The tree level (dashed) and one-loop level (solid) $\chi_1^0$
and $\chi_1^+$ masses vs. $M_2$.
The region to the left of the vertical dotted line at
$M_2$=141 (161) GeV is ruled out at one-loop (tree) level.
(b) Contours of the light chargino mass at tree and one-loop level. The
contours are labeled in GeV.}
\end{figure}

\noindent correction in this region is nearly
independent of the top quark mass and $\tan\beta$. In the region
$|\mu|>|M_2|$ the heaviest chargino and neutralino is dominantly
Higgsino and the correction is proportional to the
top quark mass and decreases with increasing $\tan\beta$. For example, in
Fig.(3a) in the region $|\mu|>|M_2|$ the contour
$\Delta M_{\chi_2^+}$=14 GeV near $|\mu|=430$ GeV
increases to $\Delta M_{\chi_2^+}$=26 GeV when $\tan\beta=1$. We show
the percent change in the $\chi_4^0$ mass in the $\mu,\ M_2$ plane
at $\tan\beta=4$ in Fig.(3b). The percentage change of the $\chi_2^+$ and
$\chi_4^0$ mass increases from 2\% for $|\mu|$ and $|M_2|\simeq$ 50 GeV to
6\% for $|M_2|\simeq 500$ GeV.

For the lightest neutralino ($\chi_1^0$) and chargino ($\chi_1^+$)
the percentage change in the mass is largest in the $\mu,\ M_2$ plane
along the upper right contour $M_{\chi_1^+}$=45 GeV. In this region the
$\chi_1^0$ and $\chi_1^+$ masses typically increase by 10-20\%
(8-10\%) for $\tan\beta\simeq1\ (\tan\beta\roughly{>}4$).
We illustrate this in
Figs.(4a,b). In Fig.(4a) we show the tree and one-loop level
$\chi_1^0$ and $\chi_1^+$ masses vs. $M_2$ for $\mu=100$ GeV and $\tan\beta$=1.
The LEP limit on the parameter $M_2$ shifts from 161 GeV to 141 GeV, so that
a smaller region of parameter space is ruled out after radiative corrections
are considered. Given this new limit on $M_2$
we can obtain a new limit for the mass of the lightest neutralino. In
this case, however, we find that the $\chi_1^0$ mass limit changes
by only 1 GeV as compared to its tree level
value. The contour $\chi_1^+$=45 GeV at $\tan\beta=1$ is
shown in Fig.(4b). Note that the lower left 45 GeV contour is
practically unchanged while the upper right 45 GeV contour shifts
appreciably; by 10 GeV for $\mu$
or $M_2 \simeq 200$ GeV and by 20 GeV if $\mu$ or $M_2 \simeq 100$ GeV.

In the region $2|\mu|>|M_2|,\ \chi_2^0$ is dominantly gaugino and
hence the typical 6-8\% change in the mass $M_{\chi_2^0}$ in this region is
essentially independent of $\tan\beta$. In the
region $|M_2|>2|\mu|$ the percentage change varies from 3-6\% for $\tan\beta=1$
to 2-4\% for $\tan\beta\roughly{>}4$. These same comments hold for the third
neutralino $\chi_3^0$, provided $2\mu$ and $M_2$ are interchanged.

\section{Conclusions}
We have computed corrections to the neutralino and chargino masses in the MSSM.
We find that typically the corrections are of order 6\%.
These corrections are of the order expected by
simple examination of the relevant Feynman diagrams.
The largest corrections, of order 20\% for the lightest neutralino,
occur on the boundary of the region in the $\mu,\ M_2$ plane
excluded by LEP measurements. These corrections can increase somewhat the
region of parameter space allowed by the latest data.

Here we included only quark/squark and lepton/slepton loops. We can expect
the corrections from the gauge/Higgs/gaugino/Higgsino sector to be of this
same order of magnitude. While the basic pattern of $\chi$ masses remains
unaltered by including radiative corrections, they should be included in the
extraction of the parameters $\mu$ and $M_2$ in the fortunate
circumstance that the charginos and neutralinos are discovered.

\end{document}